\documentclass[aps,prc,reprint,groupedaddress]{revtex4-1}
\usepackage[utf8]{inputenc}
\usepackage{geometry}
\usepackage{graphicx}
\usepackage{booktabs}
\usepackage[english]{babel}
\usepackage[svgnames]{xcolor} 

\usepackage{amsfonts, amsmath, amsthm, amssymb} 

\begin{document}


\title{Compact Astrophysical Objects in $f(R,T)$ gravity}


\author{P.H.R.S. Moraes$^1$}
\email{moraes.phrs@gmail.com}
\author{J.D.V. Arba\~nil$^2$}
\email{jd.arbanil@gmail.com}
\author{G.A. Carvalho$^1$}
\email{geanderson.araujo.carvalho@gmail.com}
\author{R.V. Lobato$^1$}
\email{vieira.lobato@gmail.com}
\author{E. Otoniel$^1$}
\email{edson.otoniel@gmail.com}
\author{R.M. Marinho Jr.$^1$}
\email{marinho.rubens@gmail.com}
\author{M. Malheiro$^1$}
\email{manuelmalheiro@gmail.com}
\affiliation{$^1$Departamento de F\'isica, Instituto Tecnol\'ogico de Aeron\'autica, S\~ao Jos\'e dos Campos, SP, 12228-900, Brazil,\\
$^2$Departamento de Ciencias, Universidad Privada del Norte,
Avenida Alfredo Mendiola 6062 Urbanizaci\'on Los Olivos, Lima, Peru}


\date{\today}

\begin{abstract}

In this article we study the hydrostatic equilibrium configuration of neutron stars (NSs) and strange stars (SSs), whose fluid pressure is computed from the equations of state $p=\omega\rho^{5/3}$ and $p=0.28(\rho-4{\cal B})$, respectively, with $\omega$ and ${\cal B}$ being constants and $\rho$ the energy density of the fluid. We also study white dwarfs (WDs) equilibrium configurations. We start by deriving the hydrostatic equilibrium equation for the $f(R,T)$ theory of gravity, with $R$ and $T$ standing for the Ricci scalar and trace of the energy-momentum tensor, respectively. Such an equation is a generalization of the one obtained from general relativity, and the latter can be retrieved for a certain limit of the theory. For the $f(R,T)=R+2\lambda T$ functional form, with $\lambda$ being a constant, we find that some physical properties of the stars, such as pressure, energy density, mass and radius, are affected when $\lambda$ is changed. We show that for some particular values of the constant $\lambda$, some observed objects that are not predicted by General Relativity theory of gravity can be attained. Moreover, since gravitational fields are smaller for WDs than for NSs or SSs, the scale parameter $\lambda$ used for WDs is small when compared to the values used for NSs and SSs.

\end{abstract}


\maketitle


\section{Introduction}

Modified gravity theories have been constantly proposed with the purpose of solving (or evading) the $\Lambda$CDM cosmological model shortcomings (check \cite{clifton/2012} for instance). 

It should be mentioned here that a gravity theory must be tested also in the astrophysical level \cite{astashenok/2013,staykov/2014,astashenok/2015b,astashenok/2015}. Strong gravitational fields found in relativistic stars could discriminate standard gravity from its generalizations. Observational data on neutron stars (NSs), for instance, can be used to investigate possible deviations from General Relativity (GR) as probes for modified gravity theories.

The modified theory to be investigated in this article will be the $f(R,T)$ gravity \cite{harko/2011}. Originally proposed as a generalization of the $f(R)$ gravity \cite{nojiri/2007,nojiri/2009}, the $f(R,T)$ theories assume that the gravitational part of the action still depends on a generic function of the Ricci scalar $R$, but also presents a generic dependence on $T$, the trace of the energy-momentum tensor. Such a dependence on $T$ would come from the consideration of quantum effects.

The recent discovery of massive ($\sim2M_{\odot}$) pulsars \cite{demorest_nature,antoniadis_science} is an important motivation to study the hydrostatic equilibrium of NSs in alternative gravity, since apparently those massive objects cannot be predicted within GR formalism. Still in the observational aspect, massive (super-Chandrasekhar) white dwarfs (WDs) have also been detected \cite{howell/2006,scalzo/2010}, what puts, once again, GR in check.

In this article, we will investigate the spherical equilibrium configuration of NSs, WDs and strange stars (SSs) in $f(R,T)$ theory of gravity. Here, it is worth mentioning that the SSs are, indeed, likely to exist. In \cite{moraes/2014c}, it was presented a method for probing the existence of such stars from the imminent detection of gravitational waves. Once such objects existence is proved, the fundamental state of matter at high densities will be understood as the strange quark matter state.

It is also important to remark that GR effects in WDs are non-negligible as it was recently investigated \cite{cmm/2018}.

In this article, we will show the results obtained for the hydrostatic equilibrium configurations of NSs, SSs and WDs in $f(R,T)$ gravity \cite{mam/2016,clmaomm/2017}.

\section{The $f(R,T)$ theory of gravity}\label{sec:frt}

Proposed by T. Harko et al., the $f(R,T)$ gravity \cite{harko/2011} assumes the gravitational part of the action depends on a generic function of $R$ and $T$, the Ricci scalar and the trace of the energy-momentum tensor $T_{\mu\nu}$, respectively. By assuming a matter lagrangian density $\mathcal{L}_m$, the total action reads
\begin{equation}\label{frt1}
S=\frac{1}{16\pi}\int d^{4}xf(R,T)\sqrt{-g}+\int d^{4}x\mathcal{L}_m\sqrt{-g}.
\end{equation} 
In (\ref{frt1}), $f(R,T)$ is the generic function of $R$ and $T$, and $g$ is the determinant of the metric tensor $g_{\mu\nu}$.

By varying (\ref{frt1}) with respect to the metric $g_{\mu\nu}$, one obtains the following FEs:
\begin{eqnarray}\label{frt2}
&&f_R(R,T)R_{\mu\nu}-\frac{1}{2}f(R,T)g_{\mu\nu}\nonumber\\&&+(g_{\mu\nu}\Box-\nabla_\mu\nabla_\nu)f_R(R,T)=8\pi T_{\mu\nu}\nonumber\\&&-f_T(R,T)T_{\mu\nu}-f_T(R,T)\Theta_{\mu\nu},
\end{eqnarray}
in which $f_R(R,T)\equiv\partial f(R,T)/\partial R$, $f_T(R,T)\equiv\partial f(R,T)/\partial T$, $\Box\equiv\partial_\mu(\sqrt{-g}g^{\mu\nu}\partial_\nu)/\sqrt{-g}$, $R_{\mu\nu}$ represents the Ricci tensor, $\nabla_\mu$ the covariant derivative with respect to the symmetric connection associated to $g_{\mu\nu}$, $\Theta_{\mu\nu}\equiv g^{\alpha\beta}\delta T_{\alpha\beta}/\delta g^{\mu\nu}$ and $T_{\mu\nu}=g_{\mu\nu}\mathcal{L}_m-2\partial\mathcal{L}_m/\partial g^{\mu\nu}$.

Taking into account the covariant divergence of (\ref{frt2}) yields
\begin{eqnarray}\label{frt3}
&&\nabla^{\mu}T_{\mu\nu}=\frac{f_T(R,T)}{8\pi -f_T(R,T)}[(T_{\mu\nu}+\Theta_{\mu\nu})\times \nonumber \\&&\nabla^{\mu}\ln f_T(R,T)
+\nabla^{\mu}\Theta_{\mu\nu}-(1/2)g_{\mu\nu}\nabla^{\mu}T].
\end{eqnarray}

We will assume the energy-momentum tensor of a perfect fluid, i.e., $T_{\mu\nu}=(\rho+p)u_\mu u_\nu-pg_{\mu\nu}$, with $\rho$ and $p$ respectively representing the energy density and pressure of the fluid and $u_\mu$ being the four-velocity tensor, which satisfies the conditions $u_\mu u^{\mu}=1$ and $u^\mu\nabla_\nu u_\mu=0$. We have, then, $\mathcal{L}_m=-p$ and $\Theta_{\mu\nu}=-2T_{\mu\nu}-pg_{\mu\nu}$. 

For the functional form of the $f(R,T)$ function above, note that originally suggested by T. Harko et al. in \cite{harko/2011}, the form $f(R,T)=R+2\lambda T$, with $\lambda$ being a constant, has been extensively used to obtain $f(R,T)$ cosmological solutions (check \cite{moraes/2015,moraes/2014b,farasat_shamir/2015} and references therein). The $f(R,T)$ gravity authors themselves have derived in \cite{harko/2011} a scale factor which describes an accelerated expansion from such an $f(R,T)$ form. Here, we propose $f(R,T)=R+2\lambda T$ to derive the $f(R,T)$ TOV equations. 

The substitution of $f(R,T)=R+2\lambda T$ in Eq.(\ref{frt2}) yields \cite{moraes/2015,moraes/2014b}
\begin{equation}\label{frt4}
G_{\mu\nu}=8\pi T_{\mu\nu}+\lambda Tg_{\mu\nu}+2\lambda(T_{\mu\nu}+pg_{\mu\nu}),
\end{equation}
for which $G_{\mu\nu}$ is the usual Einstein tensor. Equation (\ref{frt4}) is essentially the Einstein's equations of gravitation with additional terms proportional to $\lambda$, which suggest that the parameter $\lambda$ should be small. 

Moreover, when $f(R,T)=R+2\lambda T$, Eq.(\ref{frt3}) reads
\begin{align}\label{frt5}
\nabla^{\mu}T_{\mu\nu}=-\frac{2\lambda}{8\pi+2\lambda}\left[\nabla^{\mu}(pg_{\mu\nu})+\frac{1}{2}g_{\mu\nu}\nabla^{\mu}T\right].
\end{align}

\section{Equations of stellar structure in $f(R,T)$ gravity}\label{sec:tovfrt}

\subsection{Hydrostatic equilibrium equation}

In order to construct the $f(R,T)$ hydrostatic equilibrium equation, we must, firstly, develop the FEs (\ref{frt4}) for a spherically symmetric metric, such as
\begin{equation}\label{tovfrt1}
ds^2=e^{\vartheta(r)}dt^2-e^{\varpi(r)}dr^2-r^2(d\theta^2+\sin^2\theta d\phi^2).
\end{equation}

For (\ref{tovfrt1}), the non-null components of the Einstein tensor read

\begin{equation}\label{tovfrt2}
G_0^{0}=\frac{e^{-\varpi}}{r^{2}}(-1+e^{\varpi}+\varpi' r),
\end{equation}
\begin{equation}\label{tovfrt3}
G_1^{1}=\frac{e^{-\varpi}}{r^{2}}(-1+e^{\varpi}-\vartheta' r),
\end{equation}
\begin{equation}\label{tovfrt4}
G_2^{2}=\frac{e^{-\varpi}}{4r}[2(\varpi'-\vartheta')-(2\vartheta''+\vartheta'^{2}-\vartheta'\varpi')r],
\end{equation}
\begin{equation}\label{tovfrt5}
G_3^{3}=G_2^{2},
\end{equation}
for which primes stand for derivations with respect to $r$.

By substituting Eqs.(\ref{tovfrt2})-(\ref{tovfrt3}) in (\ref{frt4}) yields

\begin{equation}\label{tovfrt6}
\frac{e^{-\varpi}}{r^{2}}(-1+e^{\varpi}+\varpi'r)=8\pi\rho+\lambda(3\rho-p),
\end{equation}
\begin{equation}\label{tovfrt7}
\frac{e^{-\varpi}}{r^{2}}(-1+e^{\varpi}-\vartheta'r)=-8\pi p+\lambda(\rho-3p).
\end{equation}

As usually, we introduce the quantity $m$, representing the gravitational mass within the sphere of radius $r$, such that $e^{-\varpi}=1-2m/r$. Replacing it in \eqref{tovfrt6} yields

\begin{equation}\label{mass_continuity}
m'=4\pi r^2\rho+\frac{\lambda(3\rho-p)r^2}{2}.
\end{equation}

Moreover, from the equation for the non-conservation of the energy-momentum tensor (\ref{frt5}), we obtain

\begin{equation}\label{tovfrt8}
p'+(p+\rho)\frac{\vartheta'}{2}=-\frac{\lambda}{8\pi+2\lambda}(p'-\rho').
\end{equation}
Note that in (\ref{frt4}), (\ref{frt5}), (\ref{tovfrt6}), (\ref{tovfrt7}), (\ref{mass_continuity}) and (\ref{tovfrt8}), when $\lambda=0$ the GR predictions are retrieved.

Replacing Eq.\eqref{tovfrt7} in \eqref{tovfrt8} yields a novel hydrostatic equilibrium equation:
\begin{equation}\label{tov}
p'=-(p+\rho)\frac{\left[4\pi pr+\frac{m}{r^2}-\frac{\lambda(\rho-3p)r}{2}\right]}{\left(1-\frac{2m}{r}\right)\left[1+\frac{\lambda}{8\pi+2\lambda}\left(1-\frac{d\rho}{dp}\right)\right]}.
\end{equation}
Note that by taking $\lambda=0$ in Eq.\eqref{tov} yields the standard TOV equation \cite{tolman,oppievolkoff}. We observe that the hydrostatic equilibrium configurations are obtained only when $\frac{\lambda}{8\pi+2\lambda}\left(1-\frac{d\rho}{dp}\right)<1$. It is important to mention that to derive Eq.~\eqref{tov}, we considered that the energy density depends on the pressure ($\rho=\rho(p)$).

\subsection{Boundary conditions}

The integration of equations \eqref{mass_continuity} and \eqref{tov} starts with the values in the center ($r=0$): 

\begin{equation}\label{boundary1}
m(0)=0,\,\,\,\,\rho(0)=\rho_{c},\,\,\,\,p(0)=p_{c}.
\end{equation}

The surface of the star ($r=R$) is determined when $p(R)=0$. At the surface, the interior solution connects softly with the Schwarzschild vacuum solution. The potential metrics of the interior and of the exterior line element are linked by $e^{\vartheta(R)}=1/e^{\varpi(R)}=1-2M/R$, with $M$ representing the stellar total mass.

\subsection{Equations of state}

The relation used to derive the TOV equations is known as EoS. Once defined the EoS, the coupled differential equations \eqref{mass_continuity} and \eqref{tov} can be solved for three unknown functions $m$, $p$ and $\rho$. Recall that these coupled differential equations are integrated from the center towards the surface of the object.

To analyze the equilibrium configurations of NSs and SSs in $f(R,T)$ theory of gravity, two EoS frequently used in the literature will be considered: the polytropic and the MIT bag model EoS.

Within the simplest choices, we find that the polytropic EoS is one of the most used for the study of compact stars. Following the work developed by R.F. Tooper \cite{tooper1964}, we consider that $p=\omega\rho^{5/3}$, with $\omega$ being a constant. We choose the value of $\omega$ to be $1.475\times10^{-3}\,[\rm fm^3/MeV]^{2/3}$ as in \cite{raymalheirolemoszanchin,alz-poli-qbh}.  

To describe strange quark matter, the MIT bag model will be considered. Such an EoS describes a fluid composed by up, down and strange quarks only \cite{witten1984}. It has been applied to investigate the stellar structure of compact stars, e.g., see \cite{farhi_jaffe1984,Malheiro2003}. It is given by the relation $p=a\,(\rho-4{\cal B})$. The constant $a$ is equal to $1/3$ for massless strange quarks and equal to $0.28$ for massive strange quarks, with $m_s=250\,[\rm MeV]$ \cite{stergioulas2003}. The parameter ${\cal B}$ is the bag constant. In this work, we consider $a=0.28$ and ${\cal B}=60\,[\rm MeV/fm^3]$.

The EoS which describes the fluid properties inside WDs follows the model used for complete ionized atoms embedded in a relativistic Fermi gas of electrons \cite{Chandrasekhar1931, Chandrasekhar1935}:

\begin{eqnarray}
&&p(k_F) = \frac{1}{3\pi^2\hbar^3}\int_0^{k_F}\frac{k^4}{\sqrt{k^2+m_e^2}}dk,\label{eos}
\\ 
&&\rho(k_F) =\frac{1}{\pi^2\hbar^3}\int_0^{k_F}\sqrt{k^2+m_e^2}k^2dk \nonumber\\&&+\frac{m_N\mu_e}{3\pi^2 \hbar^3}k_F^3,\label{energy} 
\end{eqnarray}
where the last term of the right hand side of Eq. (\ref{energy}) is the ions energy contribution, and $m_N$ represents the nucleon mass, $m_e$ the electron mass, $k_F$ is the Fermi momentum, $\hbar$ is the reduced Planck constant and $\mu_e=A/Z$ is the ratio between the nucleon number $A$ and the atomic number $Z$ for ions, such that in the present work we use $\mu_e=2$, valid for He, Ca, and O WDs. We neglected the lattice ion energy contribution that is small and responsible for a small reduction of the WD radius \cite{Boshkayev2013a}.

\subsection{Numerical Method}

Once the EoS to be used are defined, the stellar structure equations will be solved numerically together with the boundary conditions for different values of $\rho_c$ and $\lambda$, through the Runge-Kutta $4$th-order method. 

\section{Equilibrium configurations of neutron stars, strange stars and white dwarfs}

\begin{figure}[h!]
\centering
\includegraphics[scale=0.45]{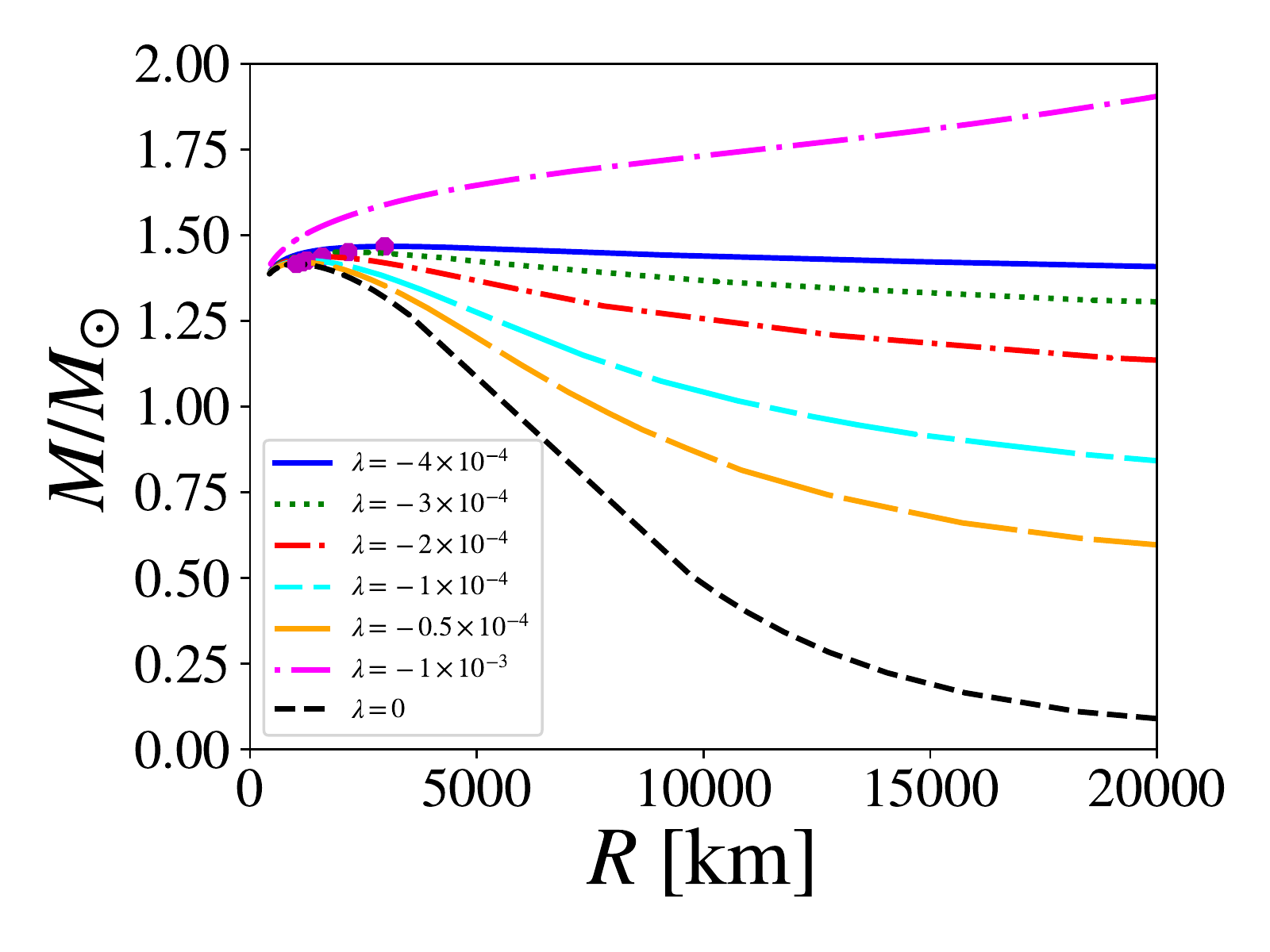}
\caption{Total mass of the white dwarfs as a function of the total radius for different values of $\lambda$. The full magenta circles indicate the maximum mass points.}
\label{fig:car}
\end{figure}

By developing the method above for the mentioned EoS with the quoted boundary conditions we obtain the figures below. In Fig.\ref{RxM_lam} we show in the upper panel the mass $\times$ radius relation for NSs and in the lower panel the relation for SSs. In Fig.\ref{fig:car} the mass $\times$ radius relation for WDs is depicted.

\begin{figure}[ht]
\centering
{\includegraphics[scale=0.24]{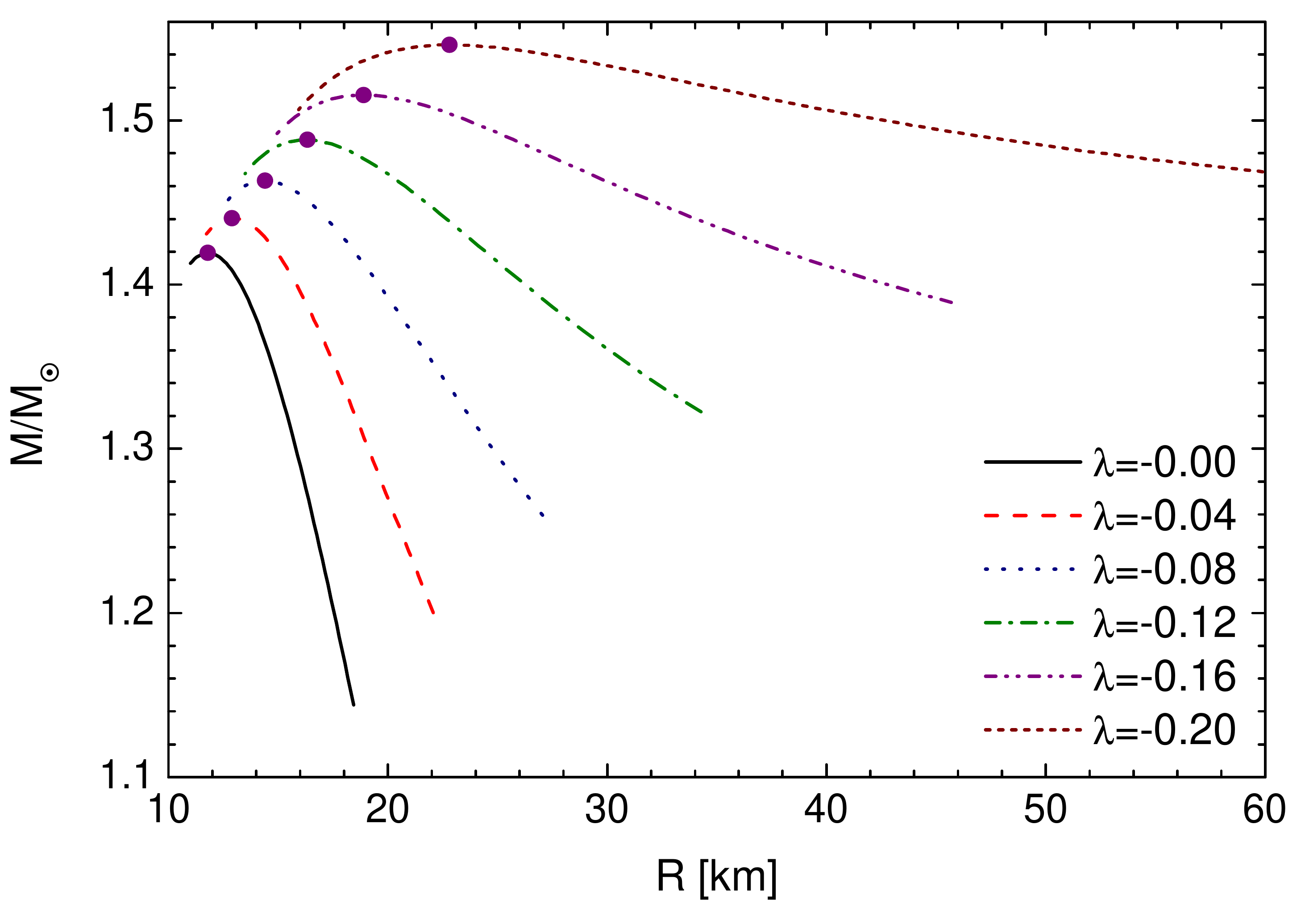}}
{\includegraphics[scale=0.35]{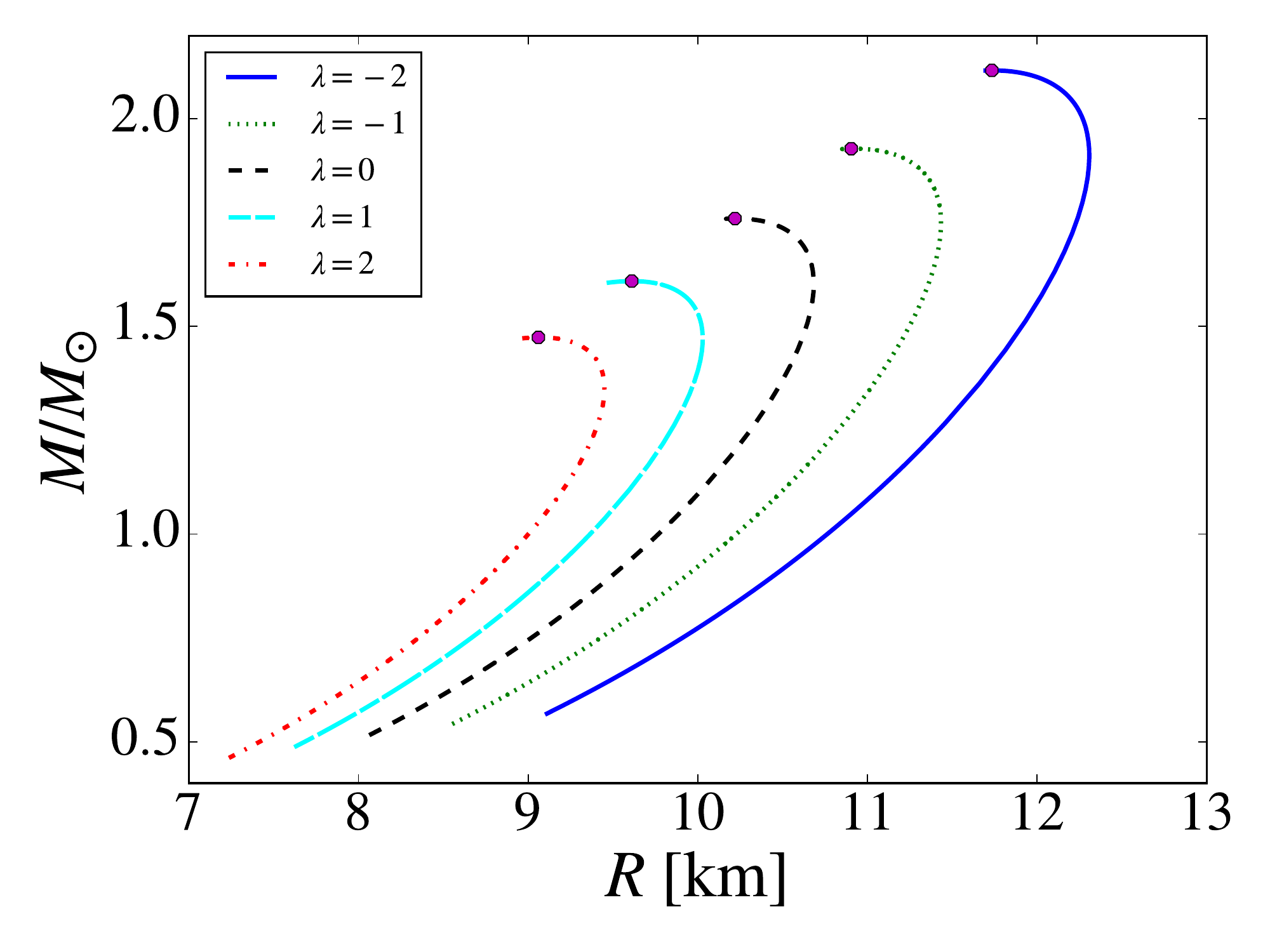}}
\caption{The total mass $M/M_{\odot}$ against the radius of the star for neutron (upper panel) and strange stars (lower panel), for five different values of $\lambda$. The maximum mass points on the curves are indicated by full circles.}
\label{RxM_lam}
\end{figure}

\section{Discussion and Conclusions}

Fig.\ref{RxM_lam} shows the behavior of the total mass, normalized in solar masses, versus the radius of the star, for different values of the parameter $\lambda$. On the upper and on the lower panels of Fig.\ref{RxM_lam} we show the results obtained for NSs and SSs, respectively. The full circles mark the maximum mass points. We can see that when we decrease the value of $\lambda$, the stars become larger and more massive. Depending on the value considered for $\lambda$, we find that both the total mass and the radius could increase between $5.834\%$ to $43.41\%$ for NSs and $20.01\%$ to $15.19\%$ for SSs.

From Figure \ref{fig:car}, we note that the masses of the WDs grow with the diminution of their total radii until attain the maximum mass point, which is represented by full magenta circles. After that, the masses decrease with the total radii. It is important to remark that the total maximum mass grows with the decrement of $\lambda$. We also mention that the curves above tend to a plateau when $\lambda$ is decreased, determining, in this way, a limit for the maximum mass of the WD, which is $\sim1.467M_{\odot}$. For smaller values of $\lambda$, $\lambda\sim -10^{-3}$, the WD stars are not stable since its mass-radius relation does not have any region respecting the stability criterion $\partial M/\partial R<0$. 

Since gravitational fields are smaller for WDs than for NSs or SSs, the scale parameter $\lambda$ used for WDs is small when compared to the values used for NSs and WDs, and also the values of $\lambda$ used for NSs are smaller than the ones used for SSs, which indicates that the more compact the star is more deviations of GR are needed and the parameter $\lambda$ may mimic a kind of chameleon mechanism, where the parameter scale depends on the density of the system \cite{brax2008}.


\end{document}